\documentclass[graybox]{svmult}

\usepackage{amsmath}
\usepackage{amssymb}

\usepackage{mathptmx}       % selects Times Roman as basic font
\usepackage{helvet}         % selects Helvetica as sans-serif font
\usepackage{courier}        % selects Courier as typewriter font
\usepackage{graphicx}        % standard LaTeX graphics tool
\usepackage[bottom]{footmisc}% places footnotes at page bottom

\usepackage{color}
\newcommand{\modif}[1]{{\color{black}{#1}}}

\renewcommand{\vec}[1]{\overrightarrow{#1}}
\begin{document}
%\preprint{APS/123-QED}

\title*{
Troubles with the radiation reaction in electrodynamics
%A time-delayed  model for radiation reaction in electrodynamics
}

%\begin{center}
\author{Sofiane Faci, Jos\'e A. Helayel-Neto and V. H. Satheeshkumar}

\institute{Sofiane Faci \at %Centro Brasileiro de Pesquisas F\'isicas, 
CBPF, Rio de Janeiro, RJ 22290-180, Brazil. \email{sofiane@cbpf.br}   
 \and
 Jos\'e A. Helayel-Neto \at %Centro Brasileiro de Pesquisas F\'isicas, 
 CBPF,  Rio de Janeiro, RJ 22290-180, Brazil. \email{helayel@cbpf.br}
 \and V. H. Satheeshkumar \at Departamento de F\'{\i}sica Te\'orica, Instituto de F\'{\i}sica, UERJ, Rio de Janeiro, Brazil;\\ %RJ 20550-900,
%and\\
Observat\'{o}rio Nacional, Rio de Janeiro,  Brazil. \email{vhsatheeshkumar@gmail.com} %RJ 20921-400,
%Departamento de F\'{\i}sica Te\'orica, Instituto de F\'{\i}sica, UERJ, Rio de Janeiro, RJ 20550-900, Brazil \email{vhsatheeshkumar@gmail.com}
}


\maketitle

\vspace{-.3cm} %if needed :o)

\abstract{
The dynamics of a radiating charge is one of the oldest unsettled problems in classical physics. The standard Lorentz-Abraham-Dirac (LAD) equation of motion is known to suffer from several pathologies and ambiguities. This paper briefly reviews these issues, and reports on a new model that fixes these difficulties in a natural way. This model is based on a hypothesis that there is an infinitesimal time delay  between action and reaction. This can be related to Feynman's regularization scheme, leading to a quasi-local QED with a natural UV cutoff,  hence without the need for renormalization as the divergences are absent. Besides leading to a pathology-free equation of motion, the new model predicts a modification of the Larmor formula that is testable with current and near future ultra-intense lasers.
}

\abstract*{
The dynamics of a radiating charge is one of the oldest unsettled problems in classical physics. The standard Lorentz-Abraham-Dirac (LAD) equation of motion is known to suffer from several pathologies and ambiguities. This paper briefly reviews these issues, and reports on a new model that fixes these difficulties in a natural way. This model is based on a hypothesis that there is an infinitesimal time delay  between action and reaction. This can be related to Feynman's regularization scheme, leading to a quasi-local QED with a natural UV cutoff,  hence without the need for renormalization as the divergences are absent. Besides leading to a pathology-free equation of motion, the new model predicts a modification of the Larmor formula that is testable with current and near future ultra-intense lasers.
}

%\pacs{03.50.-z, 03.50.De, 41.60.-m}% PACS, the Physics and Astronomy
%\keywords{Classical field theories, Classical electromagnetism, Radiation by moving charges, }%Use showkeys class option if keyword

%\tableofcontents

%%%%%%%%%%%%%%%%%%%%%%
%%%%%%%%%%%%%%%%%%%%%%
\vspace{-.1cm} %space mafia :-)
\section{Introduction}
%%%%%%%%%%%%%%%%%%%%%%
%%%%%%%%%%%%%%%%%%%%%%
The problem of electromagnetic radiation reaction %experienced by an accelerating electric charge 
goes back to the end of the nineteenth century \cite{Hammond-2010}. This history is long, rich and also particularly surprising given the simplicity of the problem at first sight. 
The standard Lorentz-invariant equation of motion of a radiating charged particle is given by the LAD equation. It is well-known that this equation is plagued by several pathologies and ambiguities. Although these have cast doubt on the foundations of classical electrodynamics,  they were long considered harmless for all practical purposes. However, the recent advances in ultra-intense laser technology \cite{DiPiazza:2011tq, Green:2013sla, Burton:2014wsa} and related sophisticated numerical simulations \cite{ji2014radiation, Dinu:2015aci} have renewed interest in this problem.
 
The LAD equation reads,
\begin{equation}\label{LAD}
m  \ddot z^\mu = F^\mu_{ext} + F^\mu_{rad},
\end{equation}
where $F^\mu_{ext}$ is the exterior Lorentz  force, and $F^\mu_{rad}$ is the radiation damping force given by,
\begin{equation}\label{LAD-force}
F^\mu_{rad} = m \epsilon (\dddot {z}^\mu + \ddot z^2\, \dot z^\mu).
\end{equation}
where $\epsilon = \frac{2 \,  e^2}{3m}$ and $\ddot z^2=\eta_{\mu\nu}\ddot z^\mu \ddot z^\nu$, with $\ddot z^\mu = \frac{d^2}{d\tau^2}z^\mu$ being the acceleration; $z^\mu=z^\mu(\tau)$ are the coordinates of the charge given as functions of the proper time $\tau$. %Note that the over dot stands for the \textbf{coordinate} time derivative. 
We use units $c=k=\hbar=1$ and the spacetime is flat with signature $%\eta_{\mu\nu}=
(+,---)$. The first term on the right hand side of the Eq. (\ref{LAD-force}) is the so-called Schott term, and the second is the Larmor term.
This is because  Larmor\modif{'s} formula for the radiated four-momentum %from a moving  particle of charge $e$ and mass $m$ 
is given by,
\begin{equation}\label{Larmor}
\delta P^\mu_{Larmor} = - m\, \epsilon \, \ddot z^2\, \dot z^\mu.
\end{equation}
Up to the current experimental precision, this formula correctly describes the observed radiated energy not only in the everyday devices like cellphones and WiFi spots, but also in the sophisticated cyclotrons and synchrotrons. 

%%%%%%%%%%%%%%%%%%%%%%%%%%%%%%%%%%%%%%%%%%%%%%%%%%% 
%%%%%%%%%%%%%%%%%%%%%%%%%%%%%%%%%%%%%%%%%%%%%%%%%%% 
\section{LAD equation: pathologies and ambiguities}
%%%%%%%%%%%%%%%%%%%%%%%%%%%%%%%%%%%%%%%%%%%%%%%%%%%
%%%%%%%%%%%%%%%%%%%%%%%%%%%%%%%%%%%%%%%%%%%%%%%%%%%

In this section, we give a brief review of the two pathologies and three ambiguities of the LAD equation.

\textbf{Self-acceleration or \textit{runaway}}. This pathology can be inferred from the non-relativistic limit of the LAD equation,
$ m  \vec{a} = \vec{f} + m\epsilon \, \dot{\vec{a}}.$ For simplicity let consider $\vec{f}=0$, the solution reads $\vec{a}(t)=\vec{a}_o \exp(\epsilon t) $, which is divergent for non-vanishing initial acceleration.
There have been several attempts to fix this pathology, among which the most notable is certainly the Landau-Lifshitz equation  \cite{Landau-Lifshitz}. This involves rewriting LAD equation (\ref{LAD}) in a perturbative way and linking  explicitly the radiation force (\ref{LAD-force}) to the external forces, this is known as order reduction. \modif{Its} non-relativistic limit reads
$m  \vec{a} = \vec{f_{ext}} + \epsilon \ \dot{\vec{f_{ext}}}+higher \ orders$. 
This equation is obviously free of runaway solutions %since the acceleration becomes trivial in the absence of external forces.
but suffers from the remaining problems of the LAD equation.
Moreover, since the perturbation parameter is given by $\dot f/f$, the Landau-Lifshitz model is limited to slowly varying external forces. 
One can also mention the similar and familiar equation of Ford and O'Connell where no divergencies appear \cite{ford1993relativistic}. 
Another attempt came from Rohrlich whose solution has the peculiarity of worsening the pre-acceleration behaviour since the charge needs to know the whole future history of the external force to adapt its acceleration \cite{rohrlich1961}. 

\textbf{Pre-acceleration.} The \modif{charge's} acceleration always precedes the external force, $m\vec{a}(t-\epsilon) \approx \vec{f}(t$), leading to causality violation.
There have been not many attempts at fixing this pathology. Since it is characterised by the infinitesimal time $\epsilon\approx 10^{-23}s$, it is believed that there could be no classical resolution. 
Quantum mechanics is required to go further eventhough it is not well suited to describe motion\footnote{It is possible to infer the equation of motion from non-relativistic QM  as a limit for averaged operators using the Ehrenfest theorem but we do not know \modif{exactly} how to describe the motion of radiating charges in this framework \cite{Moylan}.}. 
%Following this idea, several works have invoked quantum mechanics to treat the problem of moving charge. In particular, 
The first ones to implement this program were Moniz and Sharp who have used the Heisenberg picture and standard perturbative theory \cite{Moniz:1976kr}. The pre-acceleration pathology is \modif{avoided} by introducing a cutoff that corresponds to Compton scale, $\lambda = 137 \epsilon$ (\modif{recall that $c=1$}). This comes as no surprise since the cutoff is much bigger than the pathology typical scale. More recent developments include the work of Higuchi and Martin who take into consideration the full relativistic QED \cite{Higuchi:2004pr}. Unfortunately they recover \modif{the LAD equation and its associated pathologies} in the classical limit. 

%
%\begin{figure}[h]
%\sidecaption
%\includegraphics[width = 7cm]{pre-acceleration.png}
%\caption{ The acceleration appears to respond always before the external force is applied. This leads to a systematic violation of causality.}
%\label{Fig1}
%\end{figure}
%

%\section{Three ambiguities.}

%The LAD equation is also known to suffer from three ambiguities.

%\begin{itemize}
%\item
%Time-reversibility or not of the LAD equation and the radiation process
%\item
%Uniform acceleration and a possible conflict with the Equivalence Principle.
%\item 
%Energy balance paradox
%\end{itemize}

%In analogy with the usual dissipative drag force $m\vec{a}=-\lambda \vec{v}$,
\textbf{Time-reversible or not}. One might argue that the time-irreversible character of the LAD equation is obvious due to the presence of the Schott term, $\propto \dddot{z}^\mu$,  \modif{indeed}, every odd-order time-derivative of the position being irreversible. However, some authors believe that classical electrodynamics should be time reversible and sometimes prefer to rewrite the radiation force (\ref{LAD-force}) in an integro-differential form to $hide$ the Schott term \cite{jackson1999classical}. 
Rohrlich has argued that LAD equation is reversible provided that the retarded fields are replaced with advanced ones \cite{Rohrlich-1998}, % To what Zeh replied that this is equivalent to expect friction to be reversible if dissipated heat was replaced with heat focussing \cite{Zeh-1999}. 
%Subsequently, Rohrlich has defended that
but the radiation process, as a whole, is irreversible for Nature preferring retarded instead of advanced fields \cite{Rohrlich2001}. Rovelli refuted the argument stating that time reversal should also interchange cause and effect \cite{Rovelli-2004}.

\textbf{Uniform acceleration}. The problems with the LAD equation become evident when considering uniform acceleration. Instead of leading to trivial results, as one would expect, it raises more questions. Indeed it is not clear why there is no radiation \modif{damping} and the very origin of the radiated energy is mysterious  \modif{in this case} \cite{Harpaz:1998wd}.
In addition, this might give rise to a conflict with the Equivalence Principle which  \modif{locally} equates acceleration and gravitational. A free (\modif{unbound}) charge on Earth would emit energy forever, which does not \modif{seem to} happen. %(internal comment) it is better to use 'seem' to avoid stating strong and uncertain claims
This is so troublesome that Feynman claimed there could be no radiation in this case and commented that the dependence of Larmor's formula on the acceleration (instead of its variation) \textit{has led us astray} \cite{Feynman-gravity}. Since then an intense work has been devoted to this problem, see \cite{Higuchi:1996aj} and references therein. 
The accepted resolution, due to Boulware \cite{Boulware:1979qj}, asserts that a uniformly accelerated charge does radiate, but such a radiation cannot be detected by a comoving observer because it falls outside her future cone.

\textbf{Energy balance paradox}.
There is a systematic energy balance discrepancy in the LAD equation. Indeed, it is not possible to relate the work done against the radiation reaction force and the radiated energy-momentum. In other words, Larmor formula \modif{cannot be} recovered from the LAD equation. This is evident for uniform acceleration, as discussed in the previous paragraph, but is not limited to this particular case. This energy balance paradox \modif{was} recently revealed in \cite{Faci:2016glr} where 
it was also showed that the widely accepted treatment based on the bound field technique cannot fix this discrepancy. The underlying reason is that the momentum defined by Schott and later by Teitelboim is not a legitimate four-momentum for being indefinite and non-conserved.

%%%%%%%%%%%%%%%%%%%%%%%%%%%%%%%%%%%%%%
%%%%%%%%%%%%%%%%%%%%%%%%%%%%%%%%%%%%%%
\section{Time-delayed electrodynamics}
%%%%%%%%%%%%%%%%%%%%%%%%%%%%%%%%%%%%%%
%%%%%%%%%%%%%%%%%%%%%%%%%%%%%%%%%%%%%%

In this section, we discuss a recently proposed model for the motion of a classical charge which appears to fix the above difficulties \cite{Sofiane-Mario}. This model is based on the hypothesis of an infinitesimal time-delay between the action of an external electromagnetic field and the inertial reaction of elementary charges.
The time-delay is given by $\epsilon = 2 e^2/(3m)$ which is of order $10^{-23} s$ for an electron\footnote{This is comparable to the observed time delay in photoelectric effect by atoms and molecules. Indeed, the recent advances in the so-called attosecond chronoscopy have raised fundamental questions and generated an intense theoretical and experimental activity. This was predicted by Wigner \cite{Wigner:1955zz} and confirmed by direct observations. \modif{Time} scales vary around $10^{-18}s$ for small atoms and molecules. A recent proposal has demonstrated the technical possibility of reaching precision of $10^{-21}s$ by using high harmonic x-ray pulses generated with midinfrared lasers \cite{hernandez2013zeptosecond}. Hence the time shift attributed to the electron will be soon within the range of experimental capabilities.
}. 
This corresponds to $2/3$ the time that takes light to cross the classical radius of the electron.
The  \modif{infinitesimal delay parameter} $\epsilon$ \modif{ should be seen as a scalar with dimension of time (or distance if multiplied by $c$). 
Hence $\epsilon$  is  Lorentz-invariant and is thus observer-independent.}
%does not take any Lorentz boost factor when passing from one frame to another. 
\modif{Note that} no particular assumptions are required with respect to the structure, shape or size of the electron. In particular the problems related to the rigid spherical electron do not apply for this model. % \modif{Say more about the Lorentz symmetry (this is for Group 31 symposium after all!)}
The new equation of motion reads,
\begin{equation}\label{new-LAD}
f_\mu(\tau) - m\ddot z_\mu(\tau)= m \, \left| \delta \ddot z(\tau, \epsilon)^\perp_\mu \right| =  m\, \left| \sum_{n=1}^{\infty} \frac{ \epsilon^n}{n!}\, {z_\mu^{(n+2)\perp}}(\tau)\right|,
 \end{equation}
 where $\delta \ddot z(\tau, \epsilon) = \ddot z(\tau+\epsilon)-\ddot z(\tau)$, $\left| \delta \ddot z(\tau, \epsilon)^\perp_\mu \right| = s\ \delta \ddot z(\tau, \epsilon)^\perp_\mu,$ with $s=sign( \delta \ddot z(\tau, \epsilon)^\perp_o)$. 
 This guarantees that energy flux goes from the external force $f$ to the kinetic sector $\ddot z$ when the acceleration is positive and the opposite for a negative acceleration. 
 The projector on the hyperplane $\Sigma(\tau)$ orthogonal to the charge worldline (i.e. to $\dot z^\mu$) at instant $\tau$ is denoted $\perp_{\mu\nu} = \eta_{\mu\nu} - \parallel_{\mu\nu}$ with 
$ \parallel_{\mu\nu} =  \dot z_\mu\, \dot z_\nu$ being the parallel projector on the worldline. 
This is needed for consistency since $ f_\mu(\tau) \in\Sigma(\tau)$ whilst $\ddot z_\mu(\tau+\epsilon) \in\Sigma(\tau+\epsilon)$, the two hyperplanes being not parallel, except for inertial motion.
It is important to remark that $\delta \ddot z(\tau, \epsilon)$ can be equivalently replaced by $\delta f(\tau, \epsilon) =  f(\tau)-f(\tau-\epsilon)$  in this equation (and throughout the text) provided the external field is far below the Schwinger critical limit, $E_c = \frac{m^2}{e}$ (linear electrodynamics) and the frequency under the limit $\epsilon^{-1}$ (electron-positron pair creation). Both limits are far above current experimental capabilities \cite{Bulanov:2010gb}. % and therefore demanding $f^2 \ll e^2 \, E_c^2$ is a reasonable restriction.
Within these limits, and up to the first order expansion in terms of $\epsilon$, equation (\ref{new-LAD}) reduces to 
\begin{equation}\label{first-order}
m \ddot z_\mu(\tau) =f_\mu(\tau) - s\, m\epsilon \ \dddot z^\perp_\mu + o(\epsilon^2),
\end{equation}
 with now $s=sign(\dddot z^\perp_o)$.
This is the LAD equation (\ref{LAD}) when $\dddot z_o^\perp<0$, implying $s=-1$, which corresponds for example to circular motion (cyclotron and synchrotron). For $\dddot z_o^\perp>0$ the radiation force has an opposite sign in comparison with the LAD equation and this, in principle, is experimentally testable.
That is, the pre-acceleration behaviour appears only when $\dddot z_o^\perp<0$, and one has a post-acceleration for $\dddot z_o^\perp>0$. Hence pre-acceleration is not systematic and consequently not problematic. Note also that the time-irreversal character of equation (\ref{new-LAD})  is evident for even and odd high-order terms in the expansion series do not transform equally under time reversal.
As for the radiated energy-momentum, it is given by the parallel projection,
\begin{equation}\label{power}
\delta P_{rad}^\mu 
= m \, \delta \ddot z(\tau, \epsilon)^\parallel
= m\, \sum_{n=1}^{\infty} \frac{\epsilon^n}{n!}\, {z_\mu^{(n+2)\parallel}}(\tau).
\end{equation}
Like the equation of motion (\ref{LAD}), this formula is clearly time-irreversible.
The first term of the expansion corresponds to Larmor formula (\ref{Larmor}). The higher order terms are new and might drastically change  the behaviour of radiating charges in the case of rapidly changing external forces, as in high-frequency lasers experiments.
The acceleration vector being spacelike, the Larmor term is evidently positive. The odd higher order terms are  shown to be  positive \modif{in} \cite{Sofiane-Mario}. 
The even derivative terms have an indefinite sign and are time-reversible. 
However, within the validity limit of the model, the dominant term is the Larmor term and so the radiated momentum is always positive and forward oriented. In addition, performing a motion back and forth results in a null momentum coming from even terms.
Furthermore, using the identity
$ \delta \ddot z(\tau,\epsilon)^2= (\delta \ddot z(\tau,\epsilon)^\perp)^2 + (\delta \ddot z(\tau,\epsilon)^\parallel)^2,$
together with equations (\ref{new-LAD}) and (\ref{power}), defining the total momentum flux (between the instants $\tau$ and $\tau+\epsilon$) as 
$\delta P_{tot}^\mu(\tau) = m \, \delta \ddot z(\tau,\epsilon)$ 
and the internal momentum flux as $\delta P_{int}^\mu(\tau)= f^\mu(\tau)- m \ddot z^\mu(\tau)$, 
one obtains
\begin{equation}\label{conservation}
 \delta P_{tot}^2 = \delta P_{int}^2 + \delta P_{rad}^2.
\end{equation}
This formula stands for energy-momentum conservation.
It says that the total momentum,  $\delta P_{tot}$ is split into an internal flux $\delta P_{int}$, which flows between the kinetic and potential sectors, and an external flux $\delta P_{rad}$, which is dissipated.
Moreover since it involves scalar quantities, the relation (\ref{conservation}) is frame-independent.

Let us now apply the above formula for a simple and \modif{testable} example related to the ultra-high intense laser experiments. In particular, we consider a nonrelativistic electron interacting with a monochromatic plane wave laser of frequency $\omega$ and intensity $I=\frac{1}{4\pi} E_o^2$, where $\modif{E_o}$ stands for the mean value of the electric field. 
The equation of motion is given by (\ref{first-order}) with $s=-1$ (this is a cyclic motion) while the radiated power (\ref{power}) yields, 
% an infinite series that sums up into an exact expression,
%\begin{equation}
%\dot P
%= \frac{4\pi  e^2}{m \omega}I \sinh{(\epsilon \omega)}.
%%= \dot P_{Larmor} (1+ \frac{1}{6}(\epsilon \omega)^2 + o(s^4)).
%\end{equation}
%Up to the third order this yields,
\begin{equation}
\delta P_{rad}
%= \frac{4\pi  e^2}{m \omega}I \sinh{(\epsilon \omega)}
= \delta P_{Larmor} [1+ \frac{1}{6}(\epsilon \omega)^2 + o(\epsilon\omega)^4 ].
\end{equation}
where  $\delta P_{Larmor}=m\epsilon \vec{a}^2= \frac{4\pi e^2}{m}I\epsilon$ comes out of Larmor's formula (\ref{Larmor}).
Hence the new formula predicts a higher amount of radiated energy. The excess radiated power depends linearly on the intensity of the laser, $I$, and non-linearly on its frequency, $\omega$. Consequently one can remain well below the Schwinger and $\epsilon^{-1}$  limits (which limit the validity of the present model) \modif{while the experimental} conditions for testing the \modif{predicted deviation from Larmor's formula} are guaranteed, which is more easily attained by increasing the laser frequency.

%%%%%%%%%%%%%%%%%
%%%%%%%%%%%%%%%%%
\section{Summary}
%%%%%%%%%%%%%%%%%
%%%%%%%%%%%%%%%%%
\vspace{-.4cm} %space mafia :-)

In this paper, we have attempted to fix  many pathologies of the LAD % Lorentz-Abraham-Dirac (LAD) 
equation describing the motion of a radiating charge. Our model is based on a hypothesis of an infinitesimal time delay  between action and reaction. Accordingly, the force and acceleration vectors do not live on the same hyperplane orthogonal to the worldline. The orthogonal projection of the delayed force leads to the equation of motion, a discrete delay differential equation whose expansion reduces to the LAD equation at the first order and for cyclic motion. The radiated four-momentum is extracted from the parallel projection on the  worldline of the charge, which exactly reduces to Larmor formula at the first order. The higher order terms are new and experimentally testable, thanks to the recent advances in laser technology. One practical example we have outlined has precise and explicit predictions.
Finally, we would like to mention
\modif{ that the time-delay, $\epsilon$, yields a quasi-local QED exhibiting a natural UV cutoff. This might be related to Feynman's regularization scheme \cite{Feynman:1948fi} but with no need for renormalization since no divergencies need to be cured.}

%we would like to mention that the time-delay, $\epsilon$, can lead to Feynman's regularized model, that is, quasi-local QED with a natural UV cutoff.

%\vspace{-.7cm} %space mafia :-)

%%%%%%%%%%%%%%%%%%%%%%%
%%%%%%%%%%%%%%%%%%%%%%%
\begin{acknowledgement}
We thank CNPq for financial support. VHS also acknowledges financial support from FAPERJ, and is grateful to Jailson Alcaniz for the hospitality at ON. 
\end{acknowledgement}
%%%%%%%%%%%%%%%%%%%%%%%
%%%%%%%%%%%%%%%%%%%%%%%

\vspace{-.5cm} %space mafia :-)

\end{document}